\documentclass[12pt]{article}

\usepackage[round]{natbib}
\usepackage{setspace}
\usepackage{geometry}
\usepackage[section]{placeins}
\usepackage[hidelinks]{hyperref}
\usepackage{graphicx}
\usepackage{xcolor}
\usepackage{titlesec}
\usepackage[page]{appendix}
\usepackage{enumerate}
\usepackage{epigraph}
\usepackage{lipsum}

\titleformat{\subsection}
  {\bfseries\itshape}{\thesubsection.}{1em}{}
\titleformat{\subsubsection}
  {\itshape}{\thesubsubsection.}{1em}{}

\usepackage{lscape}
\usepackage{booktabs}
\usepackage{rotating}
\usepackage{multirow}
\usepackage{longtable}
\usepackage{caption}
\usepackage{subcaption}
\usepackage{float}
\usepackage{tabularx}
\usepackage{ragged2e}
\newcolumntype{Y}{>{\RaggedRight\arraybackslash}X}
\usepackage{pdflscape}
\usepackage{afterpage}

\usepackage{tikz}
\usetikzlibrary{shapes,arrows}
\usetikzlibrary{positioning}

\tikzstyle{decision} = [diamond, draw, fill=blue!20, 
    text width=4.5em, text badly centered, node distance=3cm, inner sep=0pt]
\tikzstyle{block} = [rectangle, draw, fill=blue!20, 
    text width=5em, text centered, rounded corners, minimum height=4em]
\tikzstyle{line} = [draw, -latex']
\tikzstyle{cloud} = [draw, ellipse,fill=red!20, node distance=3cm,
    minimum height=2em]

\usepackage{amsmath}
\usepackage{amssymb}
\usepackage{amsthm}
\usepackage{mathtools}
\usepackage{dsfont}

\usepackage[ruled,vlined]{algorithm2e}

\SetCommentSty{algcommstyle}

\usepackage{textcomp}
\usepackage{sourcecodepro}
\usepackage{listings}
\definecolor{commentgrey}{gray}{0.45}
\definecolor{backgray}{gray}{0.96}
\DeclareMathOperator{\Var}{Var}
\DeclareMathOperator{\Cov}{Cov}

\newcommand*{\E}{\mathbb E}

\newcommand*{\Prob}{\mathbb P}

\newcommand*{\cond}{\;\ifnum\currentgrouptype=16 \middle\fi|\;}
\newcommand{\defeq}{\vcentcolon=}

\newcommand*{\ttilde}{{\raise.17ex\hbox{$\scriptstyle\sim$}}}

\makeatletter
\newsavebox{\mybox}\newsavebox{\mysim}
\newcommand*{\distas}[1]{%
  \savebox{\mybox}{\hbox{\kern3pt$\scriptstyle#1$\kern3pt}}%
  \savebox{\mysim}{\hbox{$\sim$}}%
  \mathbin{\overset{#1}{\kern\z@\resizebox{\wd\mybox}{\ht\mysim}{$\sim$}}}%
}
\makeatother

\makeatletter
\def\moverlay{\mathpalette\mov@rlay}
\def\mov@rlay#1#2{\leavevmode\vtop{%
   \baselineskip\z@skip \lineskiplimit-\maxdimen
   \ialign{\hfil$\m@th#1##$\hfil\cr#2\crcr}}}
\newcommand*{\charfusion}[3][\mathord]{
  #1{\ifx#1\mathop\vphantom{#2}\fi\mathpalette\mov@rlay{#2\cr#3}}
  \ifx#1\mathop\expandafter\displaylimits\fi}
\makeatother

\newcommand{\suppInstPenGrpgTimet}{\ensuremath{p_{0gt}}}
\newcommand{\suppInstPenGrpgTimetHat}{\ensuremath{\hat{p}_{0gt}}}
\newcommand{\suppInstPenVec}{\ensuremath{\mathbf{p}_{0}}}
\newcommand{\suppInstPenVecHat}{\ensuremath{\hat{\mathbf{p}}_{0}}}

\newtheorem{theorem*}{Theorem}

\newtheorem{corollary*}[theorem*]{Corollary}
\newtheorem{proposition}{Proposition}[section]
\newtheorem{proposition*}[theorem*]{Proposition}

\newtheorem{lemma*}[theorem*]{Lemma}
\newtheorem{condit*}[proposition]{Condition}
\newtheorem{coroll*}[proposition]{Corollary}

\theoremstyle{definition}

\newtheorem{definition*}{Definition}

\newtheoremstyle{algodesc}{}{}{}{}{\bfseries}{.}{ }{}%
\theoremstyle{algodesc}

\usepackage{array}

\numberwithin{equation}{section}

\begin{document}

\title{An Aggregation Scheme for Increased Power}

\author{Timothy Lycurgus \& Ben B. Hansen\thanks{The authors wish to
    acknowledge valuable comments and suggestions from Brian Rowan and
    Mark White, as well as research support from the National Science
    Foundation (DMS1646108) and the Institute for Education Sciences
    (R305A120811, R305D210029).}}

\date{}

\maketitle

\abstract{

We present an aggregation scheme that
increases power in randomized controlled trials and quasi-experiments when the intervention possesses a 
robust and well-articulated theory of change. 
Longitudinal data analyzing interventions often include multiple observations on individuals, some of which may be more likely to manifest a treatment effect than others.
An intervention's theory of change provides guidance as to which of those observations are best situated to exhibit that treatment effect. 
Our \textit{p}ower-maximizing \textit{w}eighting for
\textit{r}epeated-measurements with \textit{d}elayed-effects scheme,
{PWRD} aggregation,  converts
the theory of change into a test statistic with improved asymptotic relative
efficiency, delivering tests with greater statistical power. 
We illustrate this method on an IES-funded cluster randomized trial testing the efficacy of a reading intervention designed to assist early elementary students at risk of falling behind their peers. The salient theory of change holds program benefits to be delayed and non-uniform, experienced after a student's performance stalls. In this instance, the PWRD technique's effect on power is found to be comparable to that of doubling the number of clusters in the experiment. 
}


\newpage

\tableofcontents

\newpage

\section{Introduction}
\label{sec:intro}

Many large-scale randomized controlled trials (RCTs) and high-quality
quasi-experiments are conducted only after careful vetting in national
funding competitions.  In the United States, a leading competition
for education efficacy 
studies is the Institute of Education Sciences's (IES) Education Research Grants
program, which aims to contribute to education theory
by informing stakeholders of learning interventions' costs and benefits.  
``Strong applications'' to the program are expected to detail and
justify an intervention's ``theory of change'' \citep[p.48]{ies2020}: How and why does a desired improvement
in outcomes occur as a consequence of the intervention? 
That is, what is the theory of how effects accumulate over the course of the intervention  and on which students are these benefits expected to concentrate?

This paper introduces a scheme, PWRD aggregation of effects, for
converting theories of change into statistical power for randomized controlled trials and quasi-experiments. Given an
efficacious program, a theory of change that correctly identifies where effects are likely to concentrate, and measurements indicating which students stand to benefit, this
\textit{p}ower-maximizing \textit{w}eighting for \textit{r}epeated
measurements with \textit{d}elayed effects method increases the probability
of detecting program benefits, in some cases dramatically. It is
compatible with the range of clustering accommodations and covariate
adjustment techniques that are commonly used for analysis of education RCTs.  It
maintains the canonical intention-to-treat (ITT) perspective on program
benefits.  
PWRD aggregation is applicable when there are baseline or
post-treatment measures of intervention delivery or availability, in
combination with primary outcomes measured on varying numbers of
occasions; for example, PWRD aggregation may be applied to longitudinal studies examining
the efficacy of trajectory correction interventions that provide supplemental instruction.  
The method is primarily designed to assist with hypothesis testing rather than 
with estimation yet it 
may be implemented in tandem with standard estimation techniques. 

We illustrate PWRD aggregation on an IES Education Research Grant-funded
efficacy trial of an intervention for early elementary students at
risk of falling behind in learning to read. This intervention, BURST[R]: Reading (BURST), aims to detect and correct deflections from what would otherwise be students' upward trajectory in reading ability. 
The theory of change for BURST posits this ``trajectory correction'' arises by providing targeted instruction to students whose progress has deviated from the expected course (e.g. tested below a certain benchmark). 
Thus, effects are delayed\textemdash students do not immediately
obtain an effect but must first receive
targeted remediation\textemdash and non-uniform, in that the only
students who are affected are those whose progress in reading has
slowed. As a consequence, the treatment effect will be anything but constant; if
the intervention works in the hypothesized manner, its effects will be
greatest at  follow-up times subsequent to points where student learning
would otherwise have stalled. Accordingly, beginning from estimates of the
average treatment effect (ATE) calculated separately for different cohorts of students
and occasions of follow-up, as well as information about the extent
of stalled progress at each occasion, 
PWRD aggregation combines effect estimates not only with attention to their
mutual correlations, but also with attention to their expected sizes
relative to one another.  These expectations are determined by a
carefully structured set of alternative hypotheses, which PWRD aggregation in
turn adduces from the environing theory of the intervention.

In underlying concept if not in its goals, the method relates to
instrumental variables estimation \citep{bloom1984accounting, angrist1996identification,baiocchi2014instrumental} and principal
stratification \citep{frangakis:rubi:2002,page2012,salesPane19}.
But whereas \citet{salesPane21}, for example, use principal stratification to
estimate separate effects for latent subgroups distinguished in terms
of dosage level, we marshal
related considerations to inform aggregation of effects across
manifest subgroups receiving or likely to receive differing doses.
For recent evaluation methodology using dosage information in other manners 
(e.g. to determine fidelity of implementation or to define the causal parameter of interest)
see \citet{schochet2013student} and \citet{white2019combining}.
For recent methodology proposing different weighting schemes to aggregate average treatment effect on the treated (ATT) estimates into an overall effect estimate, see \citet{callaway2021difference} and \citet{sun2021estimating}.



In this paper, we first discuss the connection of longitudinal data in
education settings to interventions with supplemental instruction to correct
stalled learning trajectories. After, we use the theory of change supporting this class of
interventions to define assumptions under which {PWRD}
aggregation will be power-maximizing. We then explicitly present the
formulation for {PWRD} aggregation weights. In Section \ref{sec:sims}, we present
a simulation study mirroring BURST design to show {PWRD} aggregation performance
in comparison with commonly used methods under various assumptions. 
In Section
\ref{sec:results}, we then illustrate how {PWRD} aggregation
compares with those same methods for BURST itself. 
Finally, in Section \ref{sec:disc}, we conclude by
summarizing how {PWRD} aggregation provides researchers with a
tool that will best help them detect an effect for interventions with
supplemental instruction.

\pagebreak

\section{Method}
\label{sec:Methods}


\subsection{Review: Comparative Studies With Repeated Measurements of the Outcome}
\label{subsec:review}


In educational settings assessing the efficacy of interventions, students frequently enter and exit studies at different points. 
For example in BURST, we examined a reading intervention on early elementary students across four years. Depending on their grade at the study's outset, the number of observations on each student varied from one to four. 
Table \ref{tab:cohort1} illustrates this phenomenon for BURST's first of four total cohorts. 

\begin{table}[htb]
\centering
\begin{tabular}{cccccc}
\toprule
            &    \textbf{Grade at Entry}  & \textbf{Year 1} & \textbf{Year 2} & \textbf{Year 3} & \textbf{Year 4} \\
                  \midrule
                  \multirow{4}{*}{\textbf{Cohort 1}}
& \textbf{3} & 3               & -               & -               & -               \\
& \textbf{2} & 2               & 3               & -               & -               \\
& \textbf{1} & 1               & 2               & 3               & -               \\
& \textbf{0} & K               & 1               & 2               & 3              \\
\bottomrule
\end{tabular}
\caption{Progression of Cohort 1 through the four years of the BURST study.}
\label{tab:cohort1}
\end{table} 

Data sources for similarly structured efficacy trials will incorporate an analogous design, with varying numbers of observations on any given participant. 
Thus, the method chosen to handle multiple observations is of great importance not only in BURST but in other longitudinal settings as well. 
The simplest outcome analysis might sidestep this debate entirely by
solely examining outcomes when students exit the study (e.g. 3rd grade
observations in BURST). For Cohort 1 in Table \ref{tab:cohort1}, this
entails using data from the diagonal and discarding the remaining
data. This method, herein termed ``exit observation'' analysis, treats
the student rather than the student-year as the unit of
analysis. Exit observation analysis typically uses models such as
\begin{equation}
\label{eq:exit_obs}
Y_{ij3} = \beta_0 + \tau Z_{ij3} + \beta X_{ij3} + \epsilon_{ij3}
\quad  \big(\E(\epsilon_{ij3})=0;\, \Var(\epsilon_{ij3})=\sigma^{2}\big), 
\end{equation}
where $Y_{ij3}$ denotes the outcome of student $i$ in school $j$ in the third grade, $X$ represents a set of demographic covariates, and $Z$ denotes the treatment status.
An example of this method may be found in \citet{simmons2008indexing}. In addition to its simplicity, exit observation analysis provides one notable benefit: an easily defined and identified overall average treatment effect, i.e. $\E[Y_{ij3}^{(Z=1)} - Y_{ij3}^{(Z=0)}]$.

However, complications emerge. According to BURST's theory of how and on whom effects will accumulate, students are more likely to benefit when they participate in the intervention for a longer period. Therefore, we are less likely to observe an effect in Cohort 1.3 than in Cohort 1.0, and treating these two groups equally may hinder a researcher's ability to detect an effect. BURST Cohort 1's experience seems to have been of this type: as seen in Table~\ref{tab:diff_means}, mean treatment-control differences in the exit observation year as compared to the entry year increase steadily from Cohort 1.2, with just 2 years of BURST, to Cohort 1.0, which enjoyed up to 4 years of BURST's supports.

\begin{table}[htb]
\centering
\begin{tabular}{crrrr}
\toprule
& \textbf{Entry} & \textbf{Entry} & \textbf{Exit} &  \multirow{2}{*}{\textbf{$\delta$}}\\
& \textbf{Grade} & \textbf{Year}  & \textbf{Year} &   \\ \midrule
\multirow{4}{*}{\textbf{Cohort 1}}
& \textbf{3} & 5.2             & 5.2   & -           \\
& \textbf{2} & -1.3            & -0.4  & 0.9         \\
& \textbf{1} & 0.3             & 3.2   & 2.9         \\
& \textbf{0} & -7.0            & 2.1   & 9.1         \\
\bottomrule    
\end{tabular}
\caption{Differences in mean reading scores between treatment and control groups for the first of four cohorts of students. The final column ($\delta$) gives the difference in these differences as calculated for the final year of participation versus the first year of participation.}
\label{tab:diff_means}
\end{table} 

In addition, exit observation analysis lacks appeal to researchers who
prefer to use all of the available data. Perhaps the easiest way to
handle repeated measurements is to fit a linear model predicting
student-year observations from independent variables identifying the
time of follow-up before estimating standard errors of these
coefficients with appropriate attention to ``clustering'' by student
or by school; in mixed modeling and general estimating equations
literature, this is known as the linear model with ``working
independence structure'' \citep{fox2015applied,laird2004analysis}. These analyses effectively attach equal weight to each student-year observation and thus we refer to them as ``flat'' weights. 
In combination with least squares, flat weighting delivers minimum-variance unbiased coefficient
estimates under the model that
\begin{equation}
\label{eq:flat}
Y_{ijk} = \beta_0 + \tau Z_{ijk} + \beta X_{ij\cdot} + \epsilon_{ijk}
\quad  \big(\E(\epsilon_{ijk})=0;\, \Var(\epsilon_{ijk})=\sigma^{2}\big), 
\end{equation}
where the disturbances $\{\epsilon_{ijk}: i,j,k\}$ \textit{are all
  independent of one another}.  The model is said only to have
``working'' independence structure because even if in
actuality the disturbances are not mutually independent, its least squares
estimates remain unbiased under Model~\ref{eq:flat}, while
clustering ensures consistency of standard errors by taking into account
heterogeneity across groups.
Model~\ref{eq:flat} differs from the
exit-observations-only model, Model~\ref{eq:exit_obs}, in
allowing multiple values of $k$ for each student $i$; in BURST,
$k$ ranges from one to four under flat weighting.
While Model~\ref{eq:flat} is general, other flat weighting identification strategies may differ. For example, one specification of flat weighting may include fixed effects for years and interactions between the treatment and year. 
An example of flat weighting may be found in \citet{meece1999changes}.

With multiple observations per student, Model~\ref{eq:flat} may be
realistic but independence of its disturbances is not; as a result,
flat weighting is inefficient. Instead of adopting this scheme,
many researchers apply mixed effects models like hierarchical linear models
\citep{raudenbush2002hierarchical} during
outcome analysis. This third option implicitly chooses a middle ground
between flat weighting and exit observation analysis. Mixed effects models
allow for some correlation between observations but not complete
correlation. In parallel with Model~\ref{eq:exit_obs} and Model~\ref{eq:flat},
we may represent the two-level mixed effects model appropriate to analysis of BURST
within the single regression equation
\begin{equation*}
\label{eq:hlm}
Y_{ijk} = \beta_0 + \tau Z_{ijk} + \beta X_{ij\cdot} + \mu_j + \epsilon_{ijk}
\quad  \big(\E(\epsilon_{ijk})=0;\, \Var(\epsilon_{ijk})=\sigma^{2}\big), 
\end{equation*}
where we adopt the same structure as with flat weighting,
including independence of $\{\epsilon_{ijk}: (i,j,k)\}$, but now
incorporate random effects $\{\mu_j: j\}$ at the school level where $\mu_j \sim N(0,\nu)$. This allows
researchers to account for unobserved heterogeneity by school. Other
formulations might incorporate an additional random effect at the
student-level. For examples of studies that apply mixed effects models, see \citet{ethington1997hierarchical},
\citet{guo2005analyzing}, and \citet{lee2000using}.

One notable drawback arises when applying the two methods utilizing more than one posttest per student. Exit observation analysis allowed us to articulate a well-defined overall average treatment effect (ATE): the expected difference in outcomes among third grade students. Using repeated measures of the outcome would seem to remove that possibility. The overall average treatment effect still represents an expected difference in outcomes between treatment and control students, but students contribute to that ATE in varying quantities depending on the length of time they participated in the study (and perhaps the intraclass correlation, or ICC).  

The presence of clustered observations, either within schools or within students, has implications beyond regression-based modeling decisions. Within-group dependence, perhaps arising due to the presence of panel data or random assignment of blocks of units, complicates standard error estimation as well. BURST data exhibit within-group dependence as a consequence of both these phenomena: treatment assignment occurred by school and we have repeated observations on multiple students. Thus, both classical and heteroskedasticity-robust standard error calculations \citep{huber1967behavior,white1980heteroskedasticity} are inappropriate. Nonetheless, dependent observations within BURST are grouped into mutually exclusive and non-overlapping clusters where every observation within the cluster received the same treatment assignment, allowing us to calculate standard errors that are robust to heterogeneity by group. For this purpose, we employ the ``cluster robust'' standard errors outlined in \citet{pustejovsky2016small}, who in turn extended the work of \citet{bell2002bias}.


\subsection{PWRD Aggregation}
\label{subsec:pwrd_present}

The three estimation methods presented in Section \ref{subsec:review} all possess certain benefits. For example, exit observation analysis allows for a well-articulated overall ATE and flat weighting allows researchers to use all of their data. Mixed effects models are particularly applicable in education settings with treatment assigned to clusters of units. 
Nonetheless, all three methods fail to take into account which observations will best allow researchers to detect a treatment effect according to the intervention's theory of change. 
In this section, we introduce an aggregation method that, similar to mixed effects models, is intermediate to flat weighting and exit observation analysis yet in contrast to those methods, leverages the intervention's theory of how effects accumulate to determine which observations are most likely to demonstrate a treatment effect.

To simplify the presentation of PWRD aggregation, we first illustrate our method on students who were in kindergarten during the first year of the study (i.e. Cohort 1.0 in Table~\ref{tab:cohort1}) for a collection of schools that implemented the intervention with some fidelity. 
These students participated in BURST for the entire study and thus, had the greatest opportunity to benefit from the intervention. 
Implementation is a post-treatment variable and we neither recommend using it in covariate adjustment nor restrict the sample to high-implementation schools when estimating treatment effects. Rather we use this subset as an example that best illustrates the intuition and process behind PWRD aggregation. 

As with the principal stratification method of \citet{salesPane21}, PWRD aggregation requires estimation of separate treatment effects for each subgroup of interest.
In \citet{salesPane21}, these are latent subgroups determined through dosage levels.  
For our method in the context of BURST, the subgroups are directly observable and refer to the cohort year of follow-up. 
Yet this too relates to dosage levels as the theory of
change suggests that these different subgroups had varying levels of exposure to the treatment: those students who had participated for longer were more likely to have received a greater dose of supplemental instruction. 
Because schools may implement the intervention differently
over time,  the method calls for separate estimates of the treatment effect for
each combination of cohort and year of follow-up. 
These covariate-adjusted treatment effect estimates for Cohort 1.0 during each year of follow-up are presented in Table~\ref{tab:ests_cohort10}. 
Note that since this is an ITT analysis, all student observations are used, not merely those exposed to the treatment.

As a departure from \citet{salesPane21} however, PWRD aggregation then serves as the tool by which we aggregate the four estimated effects in the course of hypothesis testing.  
This aggregate, similar to the two-way fixed effects difference-in-differences estimator (TWFEDD) \citep{goodman2021difference}, need not correspond to an independently meaningful average of individual effects.
However, unlike TWFEDD, PWRD aggregation does estimate its target estimand even if that estimand itself is not easily interpretable.
Therefore, this formulation simultaneously allows us to sidestep the debates reviewed in Section~\ref{subsec:review} as to how the treatment effect is best parameterized, while making use of the full, longitudinal data in a fashion best suited to detect that effect.

\begin{table}[htb]
    \centering
    \begin{tabular}{crr}
        \toprule
        \textbf{Cohort 1} & \textbf{Coef.} & \textbf{S.E.} \\ \midrule 
        \textbf{Year 1}             &  2.3           &  19.6 \\
        \textbf{Year 2}             & -9.7           &  22.6 \\
        \textbf{Year 3}             &  8.7           &  8.5  \\
        \textbf{Year 4}             & 12.8           &  10.9 \\
        \bottomrule                                      
    \end{tabular}
    \caption{Estimated change in outcome in each year of follow-up for a subset of Cohort 1.0}
    \label{tab:ests_cohort10}
\end{table}

PWRD aggregation is particularly beneficial in terms of power versus extant methods for analysis of trajectory correction interventions.
In these interventions, students only receive the treatment once their performance stalls, resulting in effects that are scattered and delayed rather than concentrated and instantaneous.  
Prior to this occurrence, students receive the same instruction they otherwise would have received if no intervention took place. 
As a consequence, the theory of change entails the exclusion
restriction \citep{angrist1996identification} that students only obtain an effect once they have received the supplemental instruction. 
The longer an individual has participated in an
intervention of this nature, the greater the likelihood of their
having become eligible to benefit from it, but prior to that occurrence, they are ``excluded'' from benefitting from the intervention.

\begin{table}[htb]
\centering
\begin{tabular}{cr}
\toprule
\multicolumn{1}{c}{\textbf{Years in BURST}} & \multicolumn{1}{c}{\textbf{Tested In}} \\ \midrule
1                                              & 66.8\%                                       \\
2                                              & 75.4\%                                       \\
3                                              & 76.7\%                                       \\
4                                              & 79.3\% \\
\bottomrule                                      
\end{tabular}
\caption{The proportion of students in Cohort 1.0 who have ``tested in'' to BURST to receive supplemental instruction by how long they have participated in the study.}
\label{tab:propry_cohort10}
\end{table}

Table~\ref{tab:propry_cohort10} shows that for Cohort 1.0, student eligibility
for the BURST intervention indeed increased in step with longer
participation in the study. This holds both for those students belonging to treatment schools and for those students attending control schools.
Accordingly, the working model describing how effects accumulate posits that the expected size of the effect in cohort $g$ during year of follow-up $t$
will be proportional to the percentage of students in cohort $g$ who were eligible for supplemental instruction by $t$, i.e. proportional to 
$\suppInstPenVec \defeq \big( \suppInstPenGrpgTimet : g, t\big)$, where $\suppInstPenGrpgTimet \defeq \Prob(${An individual in cohort $g$
is eligible to receive the supplemental instruction by year of follow-up $t$}$)$. 
Thus $\suppInstPenVec$ represents the proportion of students who were
not \textit{excluded} from having been affected, in virtue of the
assumed exclusion restriction.

The expected size of the effect as estimated through 
$\suppInstPenVecHat$
is not the only consideration of PWRD aggregation. 
Define $\Delta_{gt}$ as the parameter representing the ITT effect for cohort $g$ during year of follow-up $t$, i.e.,
\[\Delta_{gt} \defeq \E(Y_{gt}^{(Z=1)} - Y_{gt}^{(Z=0)} |G = g, T = t),\]
where $\defeq$ denotes ``defined as'', $Z = 1$ denotes assignment to the treatment  and $Z = 0$ denotes
assignment to control. Suppose corresponding ITT estimators
$\{\hat{\Delta}_{gt}: g, t\}$ to have been designated.
({PWRD} aggregation is constructed under the potential outcomes
framework of \citet{rubin1974estimating},
\citet{holland1986statistics}, and \citet{splawa1990application}. Note that our unit of observation is at the student-year level rather than at the student-level.)
The estimated relative covariances among $\{\hat{\Delta}_{gt}: g, t\}$,
denoted $\hat\Sigma$, also
factor into our method, with those effect estimates that are relatively precise and
uncorrelated with the other estimates receiving greater weight. 
This relates to precision weighting where, for example, estimates with smaller variances also receive greater weight \citep{raudenbush2002hierarchical}. 
Unlike precision weighting, however, PWRD aggregation is able to account for the correlations among cohort-year ATE estimates as well, which are often substantial. 

{PWRD} aggregation calculates a power-optimizing weighted
combination of cohort/year of follow-up ITT estimates --- an
aggregate
\begin{equation}
    \label{eq:agg_stat}
    \hat{\Delta}_{agg} \defeq \sum_{g, t} \omega_{gt}\hat{\Delta}_{gt}, 
\end{equation}
with specially chosen weights $\omega$ ($\omega_{gt}\geq
0$,  all $g, t$; $\sum_{g, t} \omega_{gt} = 1$).
To find the specific $\omega$ that maximizes power to detect an effect, we first make multiple assumptions about the nature of the treatment, 
given the theory of the intervention is correct:
\begin{condit*}
\label{cond:homo}
Individuals who receive supplemental instruction as a result of the intervention at time $j$ receive an effect $\tau \geq 0$ at some point between $j$ and $t_i$, where $t_i$ denotes the time at which individual $i$ exits the study. Individuals who do not receive supplemental instruction are unaffected. 
\end{condit*}
\begin{condit*}
\label{cond:retain}
Effect $\tau$ received by individual $i$ at time $j$ is retained by individual $i$ in full throughout the duration of the study, i.e. from $[j,t_i]$.
\end{condit*}

The second portion of Condition \ref{cond:homo} is an extension of the Stable Unit Treatment Value Assumption (SUTVA) \citep{rubin1980randomization}. 
Briefly, SUTVA states that the treatment received by one individual will not affect the potential outcomes of other individuals in the study. 
With respect to BURST, we argue this implies individuals testing into the intervention to receive targeted remediation will not affect the potential outcomes of individuals in the treatment who remain in the classroom without any supplemental instruction. 
This corresponds to a situation where there is no interference across individuals \citep{sobel2006randomized}. 
Effectively, Conditions~\ref{cond:homo} and \ref{cond:retain} amount to assuming that the effect for cohort $g$ during year of follow-up $t$ is proportional to the share of the cohort non-excluded by time $t$. 


From these conditions, we now construct {PWRD} aggregation (formally presented in Proposition~\ref{prop:main} in Appendix~\ref{sec:app_deriv}).
Take the null hypothesis that there is no effect of the intervention compared to an alternative hypothesis that there is an effect and that effect is proportional to the dosage received:
\begin{equation}
    \label{eq:hypo_test}
    \begin{split}
        & H_0: \Delta = 0 \\
        & H_a: \Delta = \eta \suppInstPenVec, \eta > 0
    \end{split}
\end{equation}
Now take test statistics centered around the aggregation of separate cohort-year ATEs, i.e. $\hat\Delta_{agg} = \sum_{g, t} \omega_{gt}\hat{\Delta}_{gt}$.
Given the above conditions hold, the asymptotic relative efficiency and thus, the power of 
these test statistics will be maximized using weights of the following form:
\begin{equation*} 
    \label{eq:opt_omega}
    \omega = (\Sigma^{-1}\suppInstPenVec)_+ \bigg/\sum_j (\Sigma^{-1}\suppInstPenVec)_{+_j},
\end{equation*}
where 
$(\Sigma^{-1}\suppInstPenVec)_{+}$ denotes the element-wise maximum of $(\Sigma^{-1}\suppInstPenVec)$ and $\mathbf{0}$, and $(\cdot)_{+_j}$ denotes the $j$th element of $(\cdot)_{+}$ such that $\omega'\mathbf{1} = 1$.
Note that Conditions~\ref{cond:homo} and \ref{cond:retain} are not required for the validity of the hypothesis test, but are necessary for finding an efficient estimator of $\Delta_{agg}$.

In sum, so long as the effect is proportional to the share of non-excluded observations, the
``signal-to-noise'' ratio of test statistics centered around $\hat\Delta_{agg}$ will be maximized by weights proportional
both to the expected sizes of cohort-year effects,
$\suppInstPenVec$,  and also the relative precisions of estimated cohort-year effects, $\Sigma$: $\omega \propto \Sigma^{-1}\suppInstPenVec$.
Note that any test statistic of the form:
\begin{equation}
\label{eq:test_stat}
\frac{\sum_{g, t}\omega_{gt}{\hat\Delta}_{gt} -  \sum_{g,t} \omega_{gt} \delta_{0gt}}{\hat{v}^{1/2}},
\end{equation}
such as the t-statistic combining estimates $\hat\Delta_{gt}$ with fixed weights $\omega_{gt}$,
will be covered by Proposition~\ref{prop:main}.

Using PWRD aggregation weights in place of some other set of weights provides test statistics with greater asymptotic relative efficiency. 
Improving relative efficiency by 20\% corresponds with a 20\% reduction in the sample size required to achieve the same level of power \citep{van2000asymptotic}. 
Thus, test statistics incorporating {PWRD} aggregation weights will provide researchers with a greater opportunity to detect an effect of the intervention when the working model of how effects accumulate holds. 
For a formal description of asymptotic relative efficiency with respect to PWRD aggregation, see Appendix~\ref{sec:app_deriv}.
While designed to assist with hypothesis testing, the method may be used in tandem with a different approach to ITT estimation like flat weighting or exit observation analysis. 
Alternatively, the researcher may forgo ITT estimation entirely and instead
present an instrumental variables estimate of a local ATE
that examines average effects across non-excluded cohorts.

In general terms, we derive these weights by selecting $\omega$ to maximize an approximation of the expected value of Equation~\ref{eq:agg_stat} known as the ``test slope''. 
After setting this term equal to zero and simplifying through a grouping of scalar quantities, we obtain {PWRD} aggregation weights. We additionally add a constraint to ensure that our aggregation weights are non-negative.
For a proof, see Appendix \ref{sec:app_deriv}. 

While we have presented PWRD aggregation using a one-sided test, there is nothing to prevent researchers from applying a two-sided test. 
In fact, if the magnitude of the negative effect increases with greater exposure, then PWRD aggregation with a two-sided test would once again increase power to detect that negative effect. 

\subsubsection{PWRD Aggregation in the BURST Evaluation}%
\label{subsubsec:pwrd_imp}

In order to implement PWRD aggregation, researchers first require estimates of
$\suppInstPenVec$ and $\Sigma$ to formulate the aggregation weights $\hat\omega$. In addition to contributing to $\hat\omega$, $\hat\Sigma$ assists in calculation of the standard error for $\hat\Delta_{agg}$. 

Neither $\suppInstPenVec$ nor $\Sigma$ is directly observed, but both can be estimated easily.
We estimate $\suppInstPenGrpgTimet$ through the proportion
$\suppInstPenGrpgTimetHat$ observed among students assigned to the
control. In the BURST example, this is the probability in cohort $g$ of
\textit{ever} having tested in by time $t$, 
rather than the probability of
testing in to supplemental instruction during year $t$: once a student becomes eligible for the first time, each subsequent observation for that student is deemed eligible as well. 
Thus, treatment received by a student in year $t$ does not affect
their weight in year $t+1$ or afterward; 
assuming the exclusion restriction, $\suppInstPenVecHat$ is
pre-treatment in the sense that treatment assignment does not affect it.
That is, $\suppInstPenVec$ is defined in terms of potential outcomes under the control. 

In theory, testing in to receive supplemental instruction from BURST solely occurred through \textit{Dynamic Indicators of Basic Early Literacy Skills} (DIBELS), a reading assessment administered as a part of this intervention. If a student's DIBELS score fell within a certain range, they were eligible for the intervention.
In practice, teachers may have used their own discretion when determining who received the supplemental instruction. 
Nonetheless, we estimate $\suppInstPenVec$ solely using DIBELS, as PWRD aggregation is consistent with ITT analysis. Thus, we construct PWRD aggregation weights using the proportions of students who \textit{should} have received the intervention if it was implemented with fidelity. 
That is, the level of non-excluded students within a given year of follow-up $t$ is the expected proportion of students who were eligible for supplemental instruction by $t$ as determined through DIBELS. 

Note that if we expected that treatment eligibility and thus, exposure differed between treatment and control groups (perhaps because treatment schools had greater incentive to provide DIBELS), we instead could have calculated PWRD aggregation weights using the proportion $\ensuremath{\hat{p}_{1gt}}$ observed among students assigned to the
treatment. 
Nonetheless, we had evidence to suggest this was not the case \citep{rowan2019summary}. 

Often $(\hat{\Delta}_{g,t}:g,t)$ will be estimates from a common
regression fit, in which case an accompanying estimate of the covariance of coefficient
estimates can be used to estimate $\Sigma$ in Equation \ref{eq:opt_omega}.
Our analysis of BURST used the Peters-Belson
\citeyearpar{peters1941method,belson1956technique} method and called
for a somewhat more elaborate calculation centered around
control-group residuals \citep{hansen2009attributing}.  
For a more thorough explanation, see \citet{rowan2019summary}.

To calculate the standard error, {PWRD} aggregation combines with standard techniques 
addressing complexities of study design such as block randomization and
assignment to treatment conditions by cluster, such as the school or
the classroom, rather than by the individual student. 
Simply, we scale the ``bread'' component of Huber-White sandwich estimators of the variance using a similar method as that presented by \citet{pustejovsky2016small}. 
With these cluster-robust standard errors, we are then able to conduct Wald tests to reject or accept the null hypotheses previously presented.

Covariate adjustment may be incorporated while estimating each individual $\Delta_{gt}$ either through design-based approaches outlined in \citet{lin2013agnostic}, \citet{hansen2009attributing}, or \citet{middleton2015unbiased}, or through more conventional model-based formulations. 
While not constructed around attributable effects \citep{rosenbaum2001effects}, we can extend {PWRD} aggregation into that setting with minor adjustments.

\subsection{Considerations When the Theory of How Effects Accumulate Fails}
\label{subsec:theory_fails}

When the theory of change correctly identifies which students will benefit from the intervention and at what point that will occur, {PWRD} aggregation maximizes the asymptotic relative efficiency 
of this method versus extant alternatives for the family of hypotheses
$K_{\eta}: \Delta = \eta \suppInstPenVec$. 
That is in BURST, if the treatment effect is proportional to the share of non-excluded observations, 
{PWRD} aggregation maximizes power.
But will PWRD aggregation have adverse effects on outcome analysis when the theory of the intervention does not hold and the proportionality assumption fails? 
In particular:
\begin{itemize}
    \item When there is no effect of the intervention, will PWRD aggregation lead to incorrect Type I errors?
    \item When the effect accrues in a different fashion than hypothesized by the theory of change, will PWRD aggregation yield less power than alternative methods?
\end{itemize}
To answer the first question, Appendix~\ref{sec:appendix_typei} proves
from weak technical conditions that PWRD aggregation maintains proper Type I error rates (rather than over or under-rejecting a null hypothesis of no effect). 

We address the second question both conceptually and through simulations presented in Section~\ref{sec:sims}.
The constant effects assumption behind base PWRD aggregation is rather strong, yet this is merely our ``working model'' of the treatment effect. The working model is informed by the intervention's theory of how effects accumulate. In BURST, for example, that theory posits that effects only accrue to students who receive the supplemental instruction.
The working model, however, goes beyond the theory of the intervention with the simplifying assumption of a constant effect.
So long as the working model is roughly accurate (i.e. the effect is proportional to the exposure, an assumption of the underlying theory of the intervention) we believe that PWRD aggregation will provide a benefit. 

For example, effects may increase in magnitude with greater exposure. In this scenario, PWRD aggregation will still provide gains to power because all else equal, the method emphasizes cohort-years with greater exposure which in turn have a greater opportunity to benefit from the increasing effect.
This occurs despite a violation of Condition~\ref{cond:homo}.

Instead, effects may decrease over time. Here, the working model is incorrect.
Yet in the context of BURST, PWRD aggregation may still provide a benefit if the intervention is successfully implemented. 
To illustrate, take the set of students who are eligible and benefit immediately after receiving the supplemental instruction. 
If that effect diminishes, then they would once again require supplemental instruction and thus, once again receive the benefit (assuming the intervention does, in fact, work). 
Therefore, cohort-years with greater cumulative exposure would likely have larger effects and PWRD aggregation would yield more power. 
On the other hand, if students who initially receive a benefit gradually lose that boost to their reading performance and do not benefit a second time, a scenario where PWRD aggregation may be harmful, we would argue that the intervention does not work and we should not detect an effect. The goal of BURST is to improve stalled reading abilities and if BURST is only a temporary palliative, then it has failed at achieving its aims. 

Our simulation study empirically examines the benefits and drawbacks of PWRD aggregation by comparing the power of $t$-tests based on 
Equation \ref{eq:agg_stat}'s $\hat{\Delta}_{agg}$, with weights $\omega$
as given by Equation \ref{eq:opt_omega}, to $t$-tests based on flat-weighted,
exit observation weighted or random effect-adjusted ITT
estimates. 
Our simulation study considers treatment effects of forms more and less favorable to PWRD aggregation, as well as a base scenario in which the working model is correct and  
$\Delta_{gt}$ is proportional to $\suppInstPenGrpgTimet$.

Our simulation study additionally provides an alternative to the above competitors that suggests analysts simply implement PWRD aggregation together with a standard method like flat weighting or exit observation analysis. 
Given the theory of the intervention holds, PWRD aggregation will yield substantially more power than extant alternatives; if instead effects accumulate in a manner different than that hypothesized by the theory of the intervention, the standard method will protect against a large loss of power. 
The step-down Dunnett procedure \citep{dunnett1991step,hothorn2008simultaneous} allows these two methods to be implemented simultaneously while maintaining valid Type I error rates.
Crucially, it avoids adding assumptions other than requiring a consistent estimate of the statistics' covariance. 
Furthermore, in the scenario that the two test statistics are highly correlated with one another, the step-down Dunnett procedure will provide power close to the power of any one of the correlated statistics. 

In fact, this step-down procedure need not solely use test statistics from standard PWRD aggregation and an analysis method typical to these settings. 
Researchers could instead include a third test statistic using a variant of PWRD aggregation that takes into account a different working model of the treatment effect, e.g. a working model that posits effects diminish or increase over time.  

Separately, failures of the $\Delta \propto \suppInstPenVec$ model
stemming from within-cluster interference can be studied analytically,
without need for simulations. 

\subsubsection{Addressing Within-Cluster Interference}
\label{subsubsec:interference}

We have interpreted the BURST theory of change to hold that a student's
outcomes may depend on her own treatment assignment but not that of
any other student --- that is, that the experiment was free of
\textit{interference} \citep{cox:1958, sobel2006randomized}. 
As applied to students within a school, this may be simplistic. A
school possesses finite resources, so its adopting a supplemental
instruction regime may transfer resources away from students not
receiving the supplement. In this scenario, Condition \ref{cond:homo}
no longer holds: students not targeted for a BURST supplement may
suffer an instructional detriment, with adverse effects on their learning. 

Addressing such \textit{spillover effects} within a classroom or school is an area
of active methodological research \citep{fletcher2010spillover,vanderweele2013mediation,gottfried2013spillover}, often calling for specialized
methods or other accommodations
\citep{sobel2006randomized,rosenbaum2007interference,vanderweele2013mediation,bowers2018models}. 
To address the common scenario of spillover within but not across
clusters, where clusters denote experimental units as assigned to treatment conditions,
the PWRD aggregation method applies without change. Specifically,
we may relax Condition \ref{cond:homo} in favor of the following: 

\begin{condit*}
\label{cond:relax}
Individual $i$ receiving supplemental instruction due to the intervention at time $j$ gains non-negative effect $\tau$ at some point between $j$ and $t_i$. Individuals who do not receive the supplemental instruction may experience an effect, positive or negative, so long as the overall effect of all students is positive in aggregate.
\end{condit*}

This allows for a corollary to standard PWRD aggregation (i.e. Proposition~\ref{prop:main}). The formal presentation of the corollary may be found in Appendix~\ref{subsec:app_int}.
Simply, the corollary argues that {PWRD} aggregation maintains
its advantage in the presence of spillover within clusters, so long as
the interference is compatible with a suitable adjustment of the
theory of the intervention. This is the situation arising in BURST:
a greater proportion of a school's students directly receiving the intervention
corresponds with a lower proportion of those students being at risk of
corresponding adverse spillover; its theory of change must hold that benefits
accruing to the first group exceed any detriment toward the
latter in aggregate.

The derivation of Proposition~\ref{cor:main} follows the same structure as the derivation of Proposition~\ref{prop:main} found in Appendix~\ref{sec:app_deriv}.


\section{Simulations}
\label{sec:sims}

In order to demonstrate how {PWRD} aggregation performs in comparison to exit observation analysis, flat weighting, mixed effects models, and a combination of PWRD aggregation with flat weighting through the step-down Dunnett procedure, we construct a simulation study mirroring the design of BURST. We generate student outcomes to compare statistical power across different scenarios using the following two-level model:
\begin{equation}
\label{eq:simgen}
\begin{split}
  Y_{ijk} & = \beta_0 + \beta_1\mathrm{Grade}_{ijk} + \mu_k + \epsilon_{ijk}\\
  \mu_k & = \gamma_0 + \nu_k
\end{split},
\end{equation}
with $\nu_k \sim N(0,\xi)$.
The outcome of student $i$ in year of follow-up $j$ at school $k$ is a function of the grade of the student and the random intercept of the school at which the student is enrolled, $\mu_k$. 
Note that fixed effects like race, gender, socio-economic status, and others could be added to this process, but were excluded as we have presented {PWRD} aggregation without covariate adjustment. Once we generate these outcomes, we perform the following two steps. 
First, we flag outcomes that fall below a given threshold as having tested into the intervention. Once a student tests in, all of their subsequent observations are flagged as well. The threshold changes by grade to adjust for natural improvement with age. 
Second, we impose artificial treatment effects on students within treatment-schools and find the corresponding power across iterations of this data generation.

We compare three variations of treatment effects in this simulation study. 
Under the first, all treatment observations flagged as having tested into the intervention receive some constant, positive effect $\tau$. 
Under the second, flagged treatment observations receive a constant, positive effect $\tau$ and unflagged treatment observations, i.e. individuals in the treatment who do not test into the intervention, receive a constant negative effect $-p \tau$ where $p \in (0,1]$. 
The third version of treatment effect imposes $\tau \sim N(l,2.5*l)$ for some $l$ to all treatment observations. 

To mirror BURST, we generate 32000 student-year observations across 26 pairs of schools with students divided roughly evenly across kindergarten through third grade. 
We assess the power provided by each of the models across 1000 iterations of this simulation study for each artificially imposed effect size.
Power for a given effect size is determined by calculating how often a model rejects a null hypothesis of no effect at the 5\% level out of the 1000 iterations. 
We use cluster-robust standard errors with clusters at the school level from the \texttt{clubSandwich} package in R \citep{pustejovsky2016small}.

\subsection{Simulation Results}
\label{subec:analysis}

We now present results from these simulations across the three variations of imposed treatment effect described previously.
For reference, the standard deviation of the outcome variable is 23.5 points. 
Following guidance from \citet{kraft2020interpreting}, we will refer to effect sizes less than than $0.05\sigma$ (1.2 points in our simulation study) as small, those between $0.05\sigma$  and $0.2\sigma$ (4.7 points) as moderate, and those greater than $0.2\sigma$ as large. 
Across 1,260 effect sizes on reading outcomes from 495 RCTs, the mean effect size was $0.17\sigma$ (4 points) and the 90th percentile was $0.5\sigma$ (11.8 points) \citep{kraft2020interpreting}. 
Thus, our simulation study examines these methods on effect sizes that frequently appear in reading interventions. 

\subsubsection{Effect 1}
\label{subsubsec:eff1}

\begin{figure}[htb]
  \begin{center}
  \includegraphics[scale = 0.5]{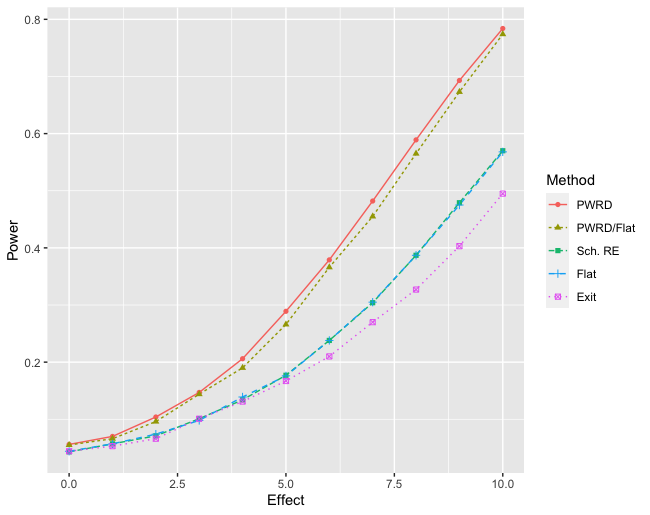}
  \end{center}
  \caption{Power for the four methods under Effect 1 across increasing effect sizes when the theory of change holds. PWRD/Flat denotes the combination of the methods through the step-down Dunnett procedure.}
    \label{fig:power_1}
\end{figure}

Figure~\ref{fig:power_1} shows the power from 1000 replications of the
synthetic experiment for each effect size across the analytical schemes mentioned above. 
The mixed effects model is specified
according to Equation~\ref{eq:simgen}, but with an independent variable
representing the treatment. It is immediately apparent that {PWRD}
aggregation outperforms the standard methods, especially for medium
effect sizes under which we observe a 35-50\% increase in power. This
is unsurprising as {PWRD} aggregation attaches greater importance to
student-year observations most likely to have received an effect from the
intervention and down-weights the remaining observations. 
Power as observed when the effect is 0 is simply the empirical size of the test; thus the left side of the plot indicates
that use of the {PWRD} method did not negatively affect Type I error
rates. 
In addition, note that while the step-down Dunnett combination of PWRD aggregation and flat weighting offers less power than PWRD aggregation, it still yields far greater power than the standard methods alone. 
Thus, this method should prove attractive both when the analyst wishes to be protected against a loss of power if the theory of change is incorrect or when the analyst wishes to both test
the null hypothesis and estimate the treatment effect using a standard approach.

\begin{figure}[htb]
  \begin{center}
  \includegraphics[scale = 0.5]{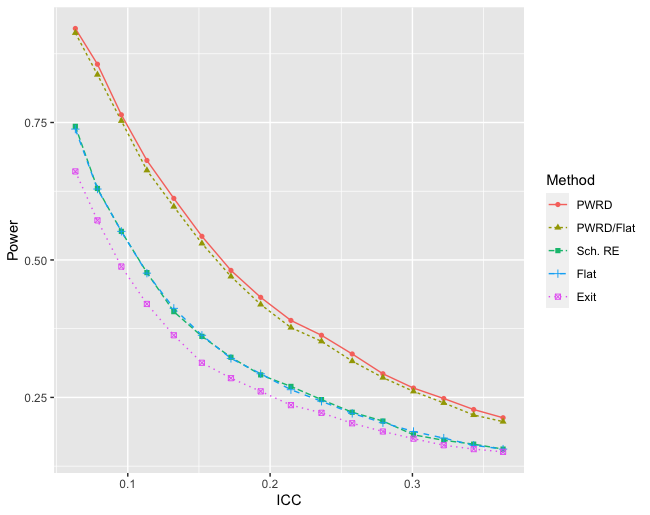}
  \end{center}
  \caption{Power for the four methods under Effect 1 with increasing intraclass correlations. PWRD/Flat denotes the combination of the methods through the step-down Dunnett procedure.}
    \label{fig:power_icc}
\end{figure}

It is natural to ask whether the gains in power present in Table
\ref{fig:power_1} hold across different levels of correlation of
observations within a school. To examine this we conducted additional simulations holding the imposed effect constant, but varying the intraclass correlation (ICC). We present these results in Figure \ref{fig:power_icc}.

In Figure \ref{fig:power_icc}, {PWRD} aggregation consistently outperforms the standard methods across ICCs that typically arise in educational settings \citep{hedges2007intraclass}. For intraclass correlations between 0.1 and 0.2, {PWRD} aggregation provides 35-45\% more power than the competitors. That gap decreases for larger ICCs, although this is at the upper range of reasonable ICC values. Furthermore, we still obtain a 35\% improvement in power. 
Lastly, the step-down Dunnett combination of PWRD aggregation and flat weighting offers substantially greater power than the standard methods do on their own. 

\subsubsection{Effect 2}
\label{subsubsec:eff2}

We now relax the assumption that students who do not receive targeted remediation through the intervention are unaffected. Instead, we impose a negative effect that is in magnitude 40$\%$ of the positive effect imposed on students who receive the supplemental instruction. 
This is a scenario where there is interference within a school, corresponding to replacing Condition \ref{cond:homo} with Condition \ref{cond:relax} and thus Proposition \ref{prop:main} with Proposition \ref{cor:main}. We chose 40$\%$ to ensure the overall effect is positive in aggregate.

\begin{figure}[htb]
  \begin{center}
  \includegraphics[scale = 0.5]{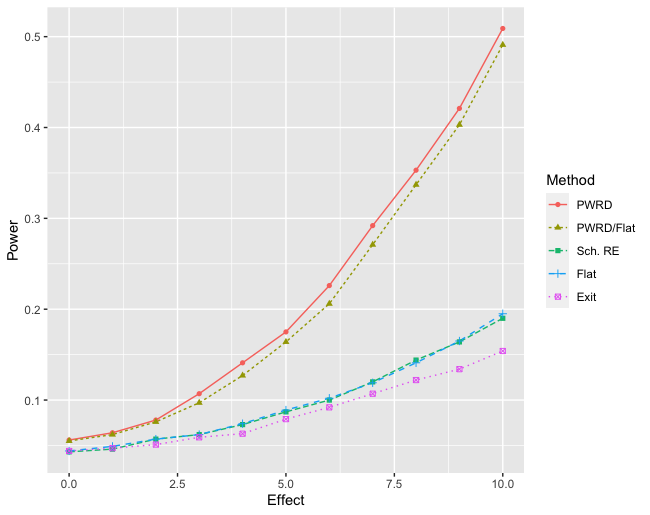}
  \end{center}
  \caption{Power for the four methods under Effect 2 across increasing effect sizes when Condition~\ref{cond:homo} does not hold and is replaced with Condition~\ref{cond:relax}. PWRD/Flat denotes the combination of the methods through the step-down Dunnett procedure.}
    \label{fig:power_2}
\end{figure}

In Figure \ref{fig:power_2}, we observe that under the relaxed assumption, {PWRD} aggregation performs even better in comparison to the traditional methods than it did under the standard assumptions. This relative gain in power is expected. We weight down effect estimates that are more likely to incorporate students with \textit{negative} effects, attaching greater importance to those more likely to have received a \textit{positive} effect. None of the other models perform a similar function and their power to detect an effect is substantially reduced as a consequence. For small effect sizes, {PWRD} aggregation increases power by roughly 30\% and this gap only widens as the effect size increases. For example, our method more than doubles the power of mixed effects models and flat weighting for large effect sizes. Once again, the step-down Dunnett combination of PWRD aggregation and flat weighting greatly outperforms the traditional alternatives although not to the extent of standard PWRD aggregation. 

\begin{figure}[htb]
  \begin{center}
  \includegraphics[scale = 0.5]{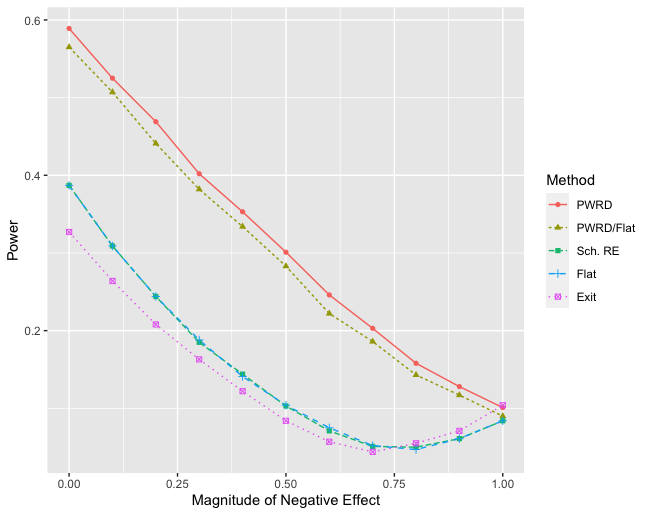}
  \end{center}
  \caption{Power for the four methods under Effect 2 with increasingly negative effects. Here we add a positive effect of size $8$ to students in the intervention and a negative effect that increases from 0\% to 100\% of the positive effect. PWRD/Flat denotes the combination of the methods through the step-down Dunnett procedure.}
  \label{fig:power_negeff}
\end{figure}

The phenomenon present in Figure \ref{fig:power_2} holds when the magnitude of the negative effect varies as well. We observe this in Figure \ref{fig:power_negeff}. Under this scenario, the size of the benefit remains constant. Instead, the adverse effect for those treatment students who do not test into the intervention varies from 0\% of the benefit to 100\% of the benefit. {PWRD} aggregation provides a persistent 15-20 percentage point advantage in power for negative effects up to 60\% of the positive effect before narrowing out. This corresponds to at least a 40\% improvement in power for all magnitudes of the negative effect; under certain circumstances, the method provides double the power. When the negative effect is equal in magnitude to the positive effect, {PWRD} aggregation no longer provides gains in power. 

\subsubsection{Effect 3}
\label{subsubsec:eff3}

\begin{figure}[htb]
  \begin{center}
  \includegraphics[scale = 0.5]{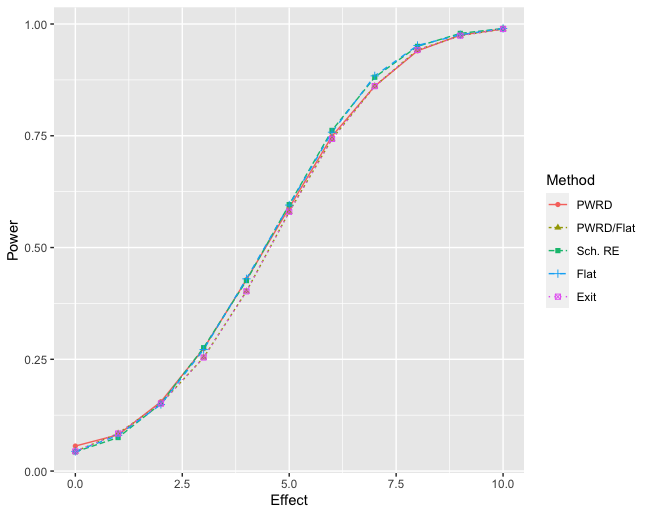}
  \end{center}
  \caption{Power for the four methods under Effect 3, i.e. across increasing effect sizes when none of Conditions~\ref{cond:homo}, \ref{cond:retain}, or \ref{cond:relax} hold. PWRD/Flat denotes the combination of the methods through the step-down Dunnett procedure.}
  \label{fig:power_3}
\end{figure}

We now examine what occurs in cases where the theory behind interventions of this sort entirely fails. 
This does not necessarily mean the intervention does not provide a benefit, just that it does not work as hypothesized by the theory of the intervention. Instead, effects may accumulate in a different fashion. 
Here, we impose an artificial treatment effect on all treatment observations such that $\tau_{ijk} \sim N(l,2.5*l)$ for $l = 1, \dots, 10$. Note that while the aggregate effect is still positive, any given student may be negatively affected. Furthermore, effects are neither stacked nor persistent across time. We present these results in Figure \ref{fig:power_3}.

We observe that while the standard methods outperform PWRD aggregation, this improvement is minimal and never exceeds 3\%. 
For example, with an imposed effect of size 5 (on the border between a moderate and large effect), the standard methods accurately reject a null hypothesis of no effect 59.6\% of the time. PWRD aggregation, on the other hand, rejects the null hypothesis 58.9\% of the time. 
For effect sizes greater than 6 (roughly $0.25\sigma)$, we are able to reject frequently under any of the three schemes.
From these simulations, it is clear that PWRD aggregation provides substantial gains in power in situations when the theory of the intervention holds. 
While an effect accumulating in a manner similar to Effect 3 may be more common than either Effect 1 or Effect 2, our simulations show only marginal decreases in its ability to reject the null hypothesis when the assumptions fail.
In this scenario, we did not require the additional protection against a failure in the intervention's theory offered by simultaneously implementing PWRD aggregation and flat weighting through the step-down Dunnett procedure. 


\section{{PWRD} Analysis Findings}
\label{sec:results}

This section presents results for BURST, both on Cohort 1.0 and on the overall randomized trial using {PWRD} aggregation and commonly applied alternative methods.  
The theory behind BURST was presented in
Section \ref{sec:Methods}. Nonetheless, its data structure merits
additional discussion to clarify analysis in this section. We utilized a
large-scale cluster randomized trial to test the efficacy of BURST, a
reading intervention designed to assist early-elementary students at
risk of falling below grade-level proficiency. The experiment was
block-randomized at the school level with 26 total blocks, 24 of which
were pairs of schools. The remaining two blocks were a triplet of
schools, in which two schools were assigned to treatment, and a
singleton. The singleton originally belonged to a pair until the
school assigned to the control attrited. Nearly every school was
matched within its school district. Across these 52 schools, we
observed 27000 unique students on 1--4 occasions each, for a total of
52000 student-year observations. 
The length of time for which each student participated in the RCT depended on the grade and year during which they entered the study. 
While we encountered some missing data, we had demographic information (race, gender, age, free lunch status, etc.) for the vast majority of students. In addition, we had DIBELS scores and end-of-year assessment scores for each student. DIBELS served as the diagnostic by which students were designated to receive targeted instruction and additionally functioned as a pre-test. The end-of-year assessments were our primary outcome of interest.
The BURST reading intervention was conducted under a University of Michigan IRB exemption.

\subsection{BURST Cohort 1.0}
\label{subsec:cohort10_results}

In this section, we begin by showing how the aggregation weights $\hat\omega$ were generated before presenting the results themselves. 
In order to calculate $\hat\omega$, we need to estimate $\suppInstPenVec$ and $\Sigma$. 
We know from Section~\ref{subsubsec:pwrd_imp} that we estimate $\suppInstPenVec$ using the proportion of control students who tested in to receive supplemental instruction for each year of follow-up. 
These values are presented in Table~\ref{tab:propry_cohort10}.
We then calculate $\Sigma$ through a grouping of control-group residuals described in greater detail in \citet{rowan2019summary}.
We then formulate:
\[\hat\omega = (\Sigma^{-1}\suppInstPenVec)_+ \bigg/\sum_j (\Sigma^{-1}\suppInstPenVec)_{+_j} = (0.25, 0, 0.32, 0.43),\]
where the weights correspond to the first through fourth years of follow-up respectively. 
Note that while more students were eligible for supplemental instruction by the second year than in the first, the relative precision of the estimate in the second year of follow-up and its mutual correlations with the other estimates were prohibitively large.
Thus, PWRD aggregation determined that outcome analysis would be best served by attaching no weight to those observations. 

We then employ a Peters-Belson \citep{peters1941method,belson1956technique} approach to estimating the average treatment effect both under standard analyses like flat weighting and mixed effects models with a random effect at the school level, and also under {PWRD} aggregation incorporating $\hat\omega$ described above. 
Briefly, Peters-Belson methods apply covariate adjustment to the control group rather than to the treatment and control simultaneously. 
That control-adjusted model is then used to predict treatment outcomes. The differences between the fitted and observed values serve to estimate the average treatment effect. 
Results are presented in Table \ref{tab:model_cohort10}.

\begin{table}[htb]
\centering
\begin{tabular}{lrrrcr}
\toprule
\textbf{Method} & \textbf{Est.} & \textbf{S.E.} & \textbf{t value} & \textbf{Sig.} & \textbf{Test Slope} \\ \midrule
Exit              &  9.88 &  9.72 &  1.02 & - & 0.082 \\
Flat              &  2.50 & 10.61 &  0.24 & - & 0.070 \\
Sch. RE           & -1.10 & 10.47 & -0.11 & - & 0.071 \\
{PWRD}            &  8.87 &  6.89 &  1.28 & - & 0.109 \\
\bottomrule                                      
\end{tabular}
\caption{BURST results on a subset of Cohort 1.0 for various methods, including PWRD aggregation.}
\label{tab:model_cohort10}
\end{table} 
None of the methods are able to detect an effect of the intervention, although PWRD aggregation provides the greatest test statistic. 
In this scenario, exit observation analysis also performs relatively well, perhaps because students in their fourth year of follow-up, i.e. in third grade, were best situated to benefit from BURST.

\subsection{BURST[R]: Reading}
\label{subsec:BURST_results}


We now conduct the same analysis described previously, yet using the complete data from BURST. 
For PWRD aggregation, we calculate separate effect estimates and aggregation weights for each cohort-year.
As with analysis on Cohort 1.0, we employ a Peters-Belson approach to covariate adjustment. Results are presented in Table~\ref{tab:BURST_res}.

\begin{table}[htb]
\centering
\begin{tabular}{lrrrcr}
\toprule
\textbf{Method} & \textbf{Est.} & \textbf{S.E.} & \textbf{t value} & \textbf{Sig.} & \textbf{Test Slope} \\ \midrule
Exit              & -1.10 &  3.25 & -0.34 & - & 0.189 \\
Flat              & -0.09 &  4.17 & -0.02 & - & 0.152 \\
Sch. RE           & -3.70 &  3.91 & -0.95 & - & 0.162 \\
{PWRD}            & -0.34 &  3.03 & -0.11 & - & 0.216 \\
\bottomrule                                      
\end{tabular}
\caption{BURST results for various methods, including PWRD aggregation.}
\label{tab:BURST_res}
\end{table} 

None of these methods detect an effect of BURST on student achievement: unfortunately, this program would appear not to have provided a benefit. 
A possible explanation for the lack of an effect is that schools possess limited resources; more students required supplemental instruction than schools had the ability to serve at levels recommended by the theory of the intervention \citep{rowan2019summary}. 
Thus, schools had to ration resources and make choices about depth of implementation versus breadth of implementation. 
These factors, along with many others, may have contributed to BURST not providing a reading benefit. 
Despite the theory of change not holding, {PWRD} aggregation still provides valid standard errors and a valid hypothesis test. 
This additionally remains the case when the intervention provides detrimental effects to students.

Nonetheless, if the theory of change were correct, the asymptotic relative efficiency of {PWRD} aggregation versus exit observation analysis, flat weighting, and mixed effects modeling was 1.30, 2.02, and 1.78 respectively. This suggests that we would have required over 15, 52, and 40 additional schools in BURST in order to achieve the same power we possessed under {PWRD} aggregation with these alternatives.


\section{Discussion}
\label{sec:disc}

The strategy of using a regression coefficient to conduct a hypothesis test is standard in settings across the social sciences. 
This approach assists in implementation of commonly used methods like exit observation analysis, flat weighting, and mixed effects models. 
Nonetheless, these conventional regressions may prove to be suboptimal in any given scenario because they fail to account for which observations are most likely to benefit from the treatment. 
In this paper, we have presented a novel method of aggregation that takes advantage of that structure by
converting the theory of how effects will accumulate in an intervention into statistical power for those interventions that provide supplemental instruction to students whose learning trajectories have stalled.
We have shown both mathematically and through a simulation study that when the working model of how effects accumulate is accurate, PWRD aggregation provides far greater power than extant alternatives.

This method is applicable in education settings, where
suitable theories of change are expected in
competitions for desirable research funding. 
We demonstrated how to extract the weights needed for PWRD aggregation from a theory of change that is likely 
to be typical of interventions providing supplemental instruction.
In it and similar circumstances, 
the method solely requires researchers possess some measure of intervention delivery or exposure in each cohort-year for either the control or treatment group.
In addition, PWRD aggregation is constructed around the theory of the intervention which is determined \textit{a priori}, so the method is consistent with pre-registration of analysis plans for increased transparency in outcome analysis. 

While {PWRD} aggregation is optimal when its supporting theory of change
holds, no benefit is gained when that theory is incorrect. Nonetheless, this scheme does not greatly hamper one's ability to detect an effect in this situation.
To further protect against any potential loss of power in settings
when the working model fails, PWRD aggregation may be used in
tandem with standard estimation techniques like exit observation
analysis or flat weighting through a step-down Dunnett procedure, as described
in Section~\ref{subsec:theory_fails}. 
In this procedure, flat weighting or exit observation analysis contributes a standard ITT estimate of the treatment effect for increased interpretability and PWRD aggregation contributes a more efficient estimator that yields greater power than the traditional method when effects accumulate in the hypothesized manner. 

We believe {PWRD} aggregation can be extended to other scenarios, both experimental and quasi-experimental, with longitudinal data and a treatment that accrues heterogeneously across observations. In each of these scenarios, similar aggregation weights can be formulated around the theory of the intervention that will maximize power.

\newpage

\bibliographystyle{apalike}
\bibliography{PWRD_Revisionv2}

\newpage

\begin{appendices}

\section{Presentation of Proposition~\ref{prop:main} and its proof}
\label{sec:app_deriv}

First take the following technical condition that simplifies the development by excluding pathological cases.
\begin{condit*}
    \label{cond:cov_conv}
    $\Cov(\hat{\Delta}) = n^{-1} \Sigma$, with $\Sigma$ a
    positive-definite symmetric matrix.
\end{condit*}

\begingroup
\setstretch{1.5}
\begin{proposition}
\label{prop:main}
Consider test statistics of the forms:
$\sum_{g,t} {\omega}_{gt}\hat{\Delta}_{gt}$, with $g$ and $t$
ranging over cohorts and times of follow-up respectively;  
$\sum_{g,t} {\omega}_{gt}\hat{\Delta}_{gt} - \sum_{g,t} \omega_{gt} \delta_{0gt}$, where $\delta_{0}$ is a vector 
of hypothesized values of $\Delta$, $\Delta \defeq (\Delta_{gt} : g,
t)$; and $\hat{v}^{-1/2}
(\sum_{g, t}{\omega}_{gt}{\hat\Delta}_{gt} - \sum_{g,t}
\omega_{gt} \delta_{0gt})$, where $\hat{v}$, perhaps an estimate
of $\Var(\sum_{g,t} {\omega}_{gt}\hat{\Delta}_{gt})$, satisfies
$n \hat{v} \rightarrow_p c > 0$.   
Consider the family of statistical hypotheses
$\{K_{\eta}: \Delta = \eta \suppInstPenVec$, $\eta \geq 0\}$. 
Under Conditions~\ref{cond:homo}, \ref{cond:retain}, and \ref{cond:cov_conv},
and for tests of $H_{0} = K_{0}$
against alternatives $K_{\eta}$, $\eta >  0$, 
asymptotic relative efficiency is maximized by 
\begin{equation} \label{eq:opt_omega}
  \omega = (\Sigma^{-1}\suppInstPenVec)_+ \bigg/\sum_j (\Sigma^{-1}\suppInstPenVec)_{+_j}.
\end{equation}
In~\eqref{eq:opt_omega}, 
$(\Sigma^{-1}\suppInstPenVec)_{+}$ denotes the element-wise maximum of $(\Sigma^{-1}\suppInstPenVec)$ and $\mathbf{0}$, and $(\cdot)_{+_j}$ denotes the $j$th element of $(\cdot)_{+}$ such that $\omega'\mathbf{1} = 1$.
\end{proposition} 
\endgroup 


Consider the parameter $\Delta_{agg} = \E(\sum_{g,t} \omega_{gt} \hat{\Delta}_{gt}) = \omega^{\prime} \Delta$ where $\Delta_{gt}$, and thus $\Delta_{agg}$, follow a proportionality assumption, i.e. $\Delta_{gt} \propto \eta\suppInstPenGrpgTimet$. 
The variance of $\omega^{\prime} \Delta$ satisfies $\Var(\sum_{g,t} \omega_{gt} \hat{\Delta}_{gt}) = \omega^{\prime} \Sigma_\Delta \omega$, where $\Sigma_\Delta$
denotes the covariance of effects across cohort-years $\{g,t\}$, and is
assumed fixed at a common value across hypotheses
$K_{\eta}$, $-\infty < \eta < \infty$. 

Now examine the test statistic $\sum_{g, t}{\omega}_{gt}\hat{\Delta}_{gt}$, the argument for the other forms being
similar. 
Our problem is to select $\omega = (\omega_{1,1},\dots,\omega_{G,T}) \geq 0$ that  maximizes the test slope of $\sum_{g, t}{\omega}_{gt}\hat{\Delta}_{gt}$ which in turn will maximize the asymptotic relative efficiency for PWRD aggregation versus alternative methods of aggregation given the theory of change is true. 
Following the definition of test slope provided in \citep[p.201]{van2000asymptotic}:
\begin{equation}
\label{eq:vdv_slope}
  h(\omega) = \frac{\Delta_{agg}'(0)}{\Cov^{1/2}_{0}(\omega^{\prime}\hat{\Delta})} = \frac{\Delta_{agg}'(0)}{\big[\omega^{\prime}\Sigma_{\Delta}\omega\big]^{1/2}},
\end{equation}
where $\Delta_{agg}'(0)$ denotes the derivative at zero of a function of the form $d \mapsto \Delta(d)$. The corresponding asymptotic relative efficiency for different $\omega$ may be represented by $\big(h(\omega_1)/h(\omega_2)\big)^2$. 
The form of the two test statistics is identical; they merely incorporate different aggregation weights $\omega$. 
Thus, it follows that finding $\omega_{opt}$, where $\omega_{opt}$ maximizes the test slope, will also maximize the asymptotic relative efficiency $\big(h(\omega_{opt})/h(\omega_{alt})\big)^2$. 
Under flat weighting, $\omega_{alt_{gt}} \defeq n_{gt}/N$, where $n_{gt}$ denotes the number of observations in cohort $g$ during year of follow-up $t$ and $N$ denotes the total number of observations. 

\subsection{Determining the Optimum $\omega_{opt}$}
\label{subsec:app_opt}

\begingroup
\setstretch{1.5}
We would like to determine which $\omega$ maximizes the test slope in (\ref{eq:vdv_slope}). 
Under the assumption that $\Delta_{g,t} \propto \eta\suppInstPenGrpgTimet$, 
then $\Delta_{gt}'(0) \propto \suppInstPenGrpgTimet$ as well. 
Thus, to determine 
which $\omega$ maximizes the test slope in (\ref{eq:vdv_slope}), we maximize the following:

\begin{equation}
\label{eq:test_slope_max}
\max_\omega \frac{\omega^{\prime}\suppInstPenVec}{\Var^{1/2}(\omega^{\prime}\hat{\Delta})}.
\end{equation}

We first transform \ref{eq:test_slope_max} logarithmically which is equivalent to maximizing $f(\omega) = \log(\omega^{\prime}\suppInstPenVec) - \frac{1}{2}\log(\Var(\omega^{\prime}\hat{\Delta}))$.
To maximize, we take the gradient of $f(\omega)$ and set the gradient equal to the zero-vector, $\mathbf{0}$, i.e. $\nabla f(\omega): \frac{\suppInstPenVec^{\prime}}{\omega^{\prime}\suppInstPenVec} - \frac{\omega^{\prime}\Sigma_\Delta}{\omega^{\prime}\Sigma_\Delta \omega} = \mathbf{0}.$
Note that both $\omega^{\prime}\suppInstPenVec$ and $\omega^{\prime}\Sigma_\Delta \omega$ are scalars, so we can rewrite this as $(\omega^{\prime}\suppInstPenVec)^{-1}\suppInstPenVec^{\prime} - (\omega^{\prime}\Sigma_\Delta \omega)^{-1}\omega^{\prime}\Sigma_\Delta = \mathbf{0}$.
We now rearrange the terms to solve for $\omega_{opt}$:
\begin{equation*}
\omega_{opt} = \Bigg(\frac{\omega^{\prime}\Sigma_\Delta \omega}{\omega^{\prime}\suppInstPenVec}\Bigg)\suppInstPenVec^{\prime} \Sigma_\Delta^{-1}.
\end{equation*}
\endgroup

\subsection{Estimation of $\omega_{opt}$}
\label{subsec:app_est}

From Slutsky's Theorem, we can then estimate $\omega_{opt}$ as follows:
\begin{equation}
  \label{eq:weights}
  \hat{\omega}_{opt} = \Bigg(\frac{\omega^{\prime}\Sigma_\Delta \omega}{\omega^{\prime}\suppInstPenVecHat}\Bigg)\suppInstPenVecHat^{\prime} \Sigma_\Delta^{-1}.
\end{equation}

\begingroup
\setstretch{1.5}
If we allow $\alpha = \Big(\frac{\omega^{\prime}\Sigma_\Delta \omega}{\omega^{\prime}\suppInstPenVecHat}\Big)$, we can then rewrite this as $\hat{\omega}_{opt} = \alpha \cdot \suppInstPenVecHat^{\prime} \Sigma_\Delta^{-1}.$ To check this simplifies, plug $\alpha \cdot \suppInstPenVecHat^\prime \Sigma_\Delta^{-1}$ back into $\omega$ in \ref{eq:weights}. We have thus uniquely specified $\hat{\omega}_{opt}$. Furthermore, in principle we can define $\hat\omega_{opt}$ only up to a constant of proportionality such that $\hat{\omega}_{opt} = \suppInstPenVecHat^\prime \Sigma_\Delta^{-1}.$ Since $\Sigma_\Delta^{-1}$ is symmetric, we can rewrite this as $\hat{\omega}_{opt} = \Sigma_\Delta^{-1} \suppInstPenVecHat.$
\endgroup

\subsection{$\omega_{opt}$ With a Non-Negativity Constraint}
\label{sec:app_nonneg}

\begingroup
\setstretch{1.5}
Previously, we wished to maximize $f(\omega) = \log(\omega^{\prime}\suppInstPenVec) - \frac{1}{2}\log(\Var(\omega^{\prime}\hat{\Delta}))$. We now add in two constraints to prevent $\omega_g < 0$. In particular, we would now like to find
$\max_\omega \log(\omega^{\prime}\suppInstPenVec) - \frac{1}{2}\log(\Var(\omega^{\prime}\hat{\Delta}))$ such that $\omega_{gt} \geq 0$ $\forall$ $\{g,t\}$ 
and $\mathbf{1}^{\prime}\omega=1$. In other words, we would like to maximize $\omega$ such that each $\omega_{gt}$ is non-negative and $\sum_{g=1}^{G} \sum_{t=1}^{T}\omega_{gt}=1$.
This is equivalent to solving: $max_{\omega}  \log(\omega^{\prime}\suppInstPenVec) - \frac{1}{2}\log(\Var(\omega^{\prime}\hat{\Delta})) - u^{\prime}\omega + v^{\prime}\omega.$
\endgroup

\begingroup
\setstretch{1.5}
We begin by looking at the KKT conditions \citep{karush1939minima,kuhn2014nonlinear}:
\begin{itemize}
  \item \textbf{Stationarity:} $(\omega^{\prime}\suppInstPenVec)^{-1}\suppInstPenVec^{\prime} - (\omega^{\prime}\Sigma_\Delta \omega)^{-1}\omega^{\prime}\Sigma_\Delta - u^{\prime} + v^{\prime} = \mathbf{0}$.

  Note: Both $(\omega^{\prime}\suppInstPenVec)^{-1}$ and $(\omega^{\prime}\Sigma_\Delta \omega)^{-1}$ are scalar random variables, so for ease we redefine them as $c_1$ and $c_2$ respectively, i.e. $c_1\suppInstPenVec^{\prime} - c_2\omega^{\prime}\Sigma_\Delta - u^{\prime} + v^{\prime} = \mathbf{0}$.

  \item \textbf{Complementary Slackness:} $u^{\prime}\omega = 0.$

  \item \textbf{Primal Feasibility:} $\omega \geq 0, \mathbf{1}^{\prime}\omega=1.$

  \item \textbf{Dual Feasibility:} $u \geq 0.$

\end{itemize}
\endgroup

To solve this, we begin by eliminating $u$, giving us $v^{\prime} - u^{\prime} = c_2\omega^{\prime}\Sigma_\Delta - c_1\suppInstPenVec^{\prime} \Rightarrow v^{\prime} \geq c_2\omega^{\prime}\Sigma_\Delta - c_1\suppInstPenVec^{\prime}$
from stationarity, and $(c_1\suppInstPenVec^{\prime} - c_2\omega^{\prime}\Sigma_\Delta + v^{\prime})\omega = 0$
from complementary slackness. After rearranging, we see that
\begin{equation*}
  \mathbf{0} \leq \omega^{\prime} \leq \frac{v^{\prime} + c_1\suppInstPenVec^{\prime}}{c_2}\Sigma_\Delta^{-1}.
\end{equation*}
From this, we then argue that $\omega_{opt}$ is maximized by the following:
\begin{equation*}
\omega_{gt} = 
\begin{cases}
  \Big(\frac{v^{\prime} + c_1\suppInstPenVec^{\prime}}{c_2}\Sigma_\Delta^{-1}\Big)_{gt} & \text{if $v_{gt} \geq -c_1\suppInstPenGrpgTimet$} \\
  0 & \text{if $v_{gt} < -c_1\suppInstPenGrpgTimet$}
\end{cases}.
\end{equation*}

In other words, $\omega_{opt}^{\prime} = (\frac{v^{\prime} + c_1\suppInstPenVec^{\prime}}{c_2}\Sigma_\Delta^{-1})_{+}$ where $\mathbf{1}^{\prime}\omega=1$. We can then estimate $\omega_{opt}$ following the same argument as in Appendix \ref{subsec:app_est}.

\subsection{PWRD Aggregation with Interference}
\label{subsec:app_int}

\begingroup
\setstretch{1.5}
\begin{proposition}
  \label{cor:main}
  Under Conditions \ref{cond:retain}, \ref{cond:relax}, and \ref{cond:cov_conv}, the following aggregation weights $\omega$ will maximize the slope of test statistics 
discussed in Proposition~\ref{prop:main}
for the family of hypothesis tests and alternative hypotheses also elaborated in Proposition \ref{prop:main}: 
  \[\omega =  (\Sigma^{-1}\suppInstPenVec)_+\bigg/\sum_j (\Sigma^{-1}\suppInstPenVec)_{+_j}.\]
\end{proposition}
\endgroup

\section{PWRD Aggregation and Type I Errors}
\label{sec:appendix_typei}

In Section~\ref{subsec:pwrd_present}, we demonstrated how {PWRD} aggregation maximizes the test slope and thus, the corresponding power for the family of hypotheses
$K_{\eta}: \Delta = \eta \suppInstPenVec$. That is, when the treatment
effect is proportional to the share of non-excluded observations, {PWRD} aggregation
maximizes power. Here, we remove that assumption and all assumptions
about the form of the treatment effect. We do require joint limiting
Normality of $\hat{\Delta}$ and a consistent estimator of its
covariance. 
\begin{condit*}
\label{cond:cov_hat_conv}
The estimator $\widehat\Cov(\hat{\Delta})$ is consistent for $\Cov(\hat{\Delta})$, in the
sense that $\lVert n\widehat\Cov(\hat{\Delta}) - \Sigma \rVert_2
\rightarrow_P 0$, where $\Sigma$ is as in Condition~\ref{cond:cov_conv}.
\end{condit*}
\begin{condit*}
\label{cond:est_mvn}
$\sqrt{n}(\hat\Delta - \Delta) \rightarrow_d N\big(\mathbf{0}, \Cov(\Delta)\big)$.
\end{condit*}
With Conditions~\ref{cond:cov_conv}, \ref{cond:cov_hat_conv} and \ref{cond:est_mvn}, we formulate a simple proposition about the distribution of the test statistic specified in Equation~\ref{eq:test_stat}.

\begingroup
\setstretch{1.5}
\begin{proposition}
  \label{prop:t_dist}
  Take fixed aggregation weights $w$. Under the null hypothesis $H_0$ and when Conditions~\ref{cond:cov_conv}, \ref{cond:cov_hat_conv}, and \ref{cond:est_mvn} hold, 
  \[\frac{\sum_{g,t} w_{gt} \hat\Delta_{gt} - \sum_{g,t} w_{gt}\delta_{0gt}}{(w'{\Cov}(\hat{\Delta})w)^{1/2}} \rightarrow_d N(0,1).\] 
\end{proposition}
\endgroup

Proposition~\ref{prop:t_dist} 
states that with a consistent estimator of the covariance and an estimator that is asymptotically multivariate normal, the test statistic specified in Equation~\ref{eq:test_stat} with fixed aggregation weights $w$ will converge to a standard multivariate normal distribution. For finite sample sizes $n$, this test statistic should approximately follow a t-distribution with $n-k$ degrees of freedom, where $k$ represents the number of estimated parameters.
Note that the denominator, $\hat{v}^{1/2}$, present in Equation~\ref{eq:test_stat} and Section~\ref{subsec:pwrd_present} at large denotes the quadratic form of estimated covariances of $\hat{\Delta}$. 
{PWRD} aggregation requires statisticians provide a covariance estimator with consistency guarantees, i.e. Condition~\ref{cond:cov_hat_conv}.

While Proposition~\ref{prop:t_dist} allows us to determine the asymptotic distribution of test statistics with the form in Equation~\ref{eq:test_stat} for fixed aggregation weights $w$, 
{PWRD} aggregation does not incorporate fixed weights.  
Rather, two components of {PWRD} aggregation, 
$\suppInstPenVecHat$ and $\hat{\Sigma}$, are random variables. 
Consequently, the aggregated statistic $\sum_{g,t}\hat{\omega}_{gt}\hat\Delta_{gt}$ 
includes an auxiliary statistic: $\hat{\omega}_{gt}$.
Addressing additional variation of this type generally requires analysis through stacked estimating equations, a technique not readily compatible with the best-in-class clustered standard error estimation of \citet{pustejovsky2016small}. 
Thus, our standard error scales the covariance between each $\hat\Delta_{gt}$ by aggregation weights $\hat{\omega}$, yet does not incorporate the covariance between each $\hat{\omega}_{gt}$. 
To address this issue, we first present a mild condition on $\suppInstPenVecHat$.
\begin{condit*}
\label{cond:rootn}
$\suppInstPenVecHat$ is root-$n$ consistent, i.e. $\lVert
\suppInstPenVecHat - \suppInstPenVec \rVert_2 = O_P(n^{-1/2})$.
\end{condit*}
As applied to the BURST study, Condition~\ref{cond:rootn} is
immediate from the Weak Law of Large Numbers. Conditions~\ref{cond:cov_conv}, \ref{cond:cov_hat_conv}, and \ref{cond:rootn} allow us to circumvent our standard error not incorporating additional variation from $\hat\omega$ through  
Proposition \ref{prop:testin_var}.

\begingroup
\setstretch{1.5}
\begin{proposition}
\label{prop:testin_var}

Consider t-statistics of the form 
\begin{equation}\label{eq:2}
  \frac{(\sum_{g,t} \hat{\omega}_{gt}\hat{\Delta}_{gt} - \sum_{g,t} \hat{\omega}_{gt}\delta_{0gt})}{(\hat{\omega}'\widehat{\Cov}(\hat{\Delta})\hat{\omega})^{1/2}},
\end{equation}
where $\hat\omega = (\widehat\Cov[\hat{\Delta}]^{-1}\suppInstPenVecHat)_+ \bigg/\sum_j
(\widehat\Cov[\hat{\Delta}]^{-1}\suppInstPenVecHat)_{+_j} \in [0,1]$ represents weights for
{PWRD} aggregation. Under Conditions~\ref{cond:cov_conv},
\ref{cond:cov_hat_conv}, and \ref{cond:rootn}, the difference between
\eqref{eq:2} and 
\begin{equation*}
  \frac{(\sum_{g,t} \omega_{gt}\hat{\Delta}_{gt} - \sum_{g,t} \omega_{gt}\delta_{0gt})}{(\omega'{\Cov}(\hat{\Delta})\omega)^{1/2}},
\end{equation*}
where $\omega = (\Sigma^{-1}\suppInstPenVec)_+ \bigg/\sum_j
(\Sigma^{-1}\suppInstPenVec)_{+_j}$,
is asymptotically negligible:
\begin{equation}
  \label{eq:t_cp}
   \Bigg[\frac{\sum_{g,t} \hat{\omega}_{gt}\hat{\Delta}_{gt} - \sum_{g,t}\hat{\omega}_{gt}\delta_{0gt}}{(\hat{\omega}'\widehat{\Cov}(\hat{\Delta})\hat{\omega})^{1/2}} - \frac{\sum_{g,t}\omega_{gt}\hat{\Delta}_{gt} - \sum_{g,t} \omega_{gt}\delta_{0gt}}{(\omega'{\Cov}(\hat{\Delta})\omega)^{1/2}}\Bigg] \rightarrow_{P} 0.
\end{equation}
\end{proposition}
\endgroup

Simply, Proposition~\ref{prop:testin_var} states that the t-statistic
centered around $\sum_{g,t} \hat{\omega}_{gt}\delta_{0gt}$, where
$\hat{\omega} = (\hat{\Sigma}^{-1}\suppInstPenVecHat)_+\bigg/\sum_j
(\hat\Sigma^{-1}\suppInstPenVecHat)_{+_j}$, and scaled by a
consistently estimated standard error will converge in
probability to the ``proto'' t-statistic appearing in
Prop.~\ref{prop:t_dist} and covered by Prop.~\ref{prop:main}, which is
centered around the parameter $\sum_{g,t}\omega_{gt}\delta_{0gt}$ and
scaled by the sampling s.d. of $\sum_{g,t} \omega_{gt}\hat\Delta_{gt}$. 
As a consequence, hypothesis tests incorporating {PWRD} aggregation will maintain proper Type I error rates. Therefore, {PWRD} aggregation provides valid hypothesis tests even when the theory of change does not hold. The proof of Proposition \ref{prop:testin_var} can be found in Appendix~\ref{sec:proof_testin_var}, the following subsection. 

\subsection{Proof of Proposition \ref{prop:testin_var}}
\label{sec:proof_testin_var}

To show Proposition \ref{prop:testin_var}, we begin by showing that
$\lVert \hat{\omega} - \omega\rVert_2 \rightarrow_{P}0$. Writing
$\hat{\Sigma} \defeq  n \widehat\Cov(\hat{\Delta})$,
Condition~\ref{cond:cov_hat_conv} says $\lVert \hat\Sigma - \Sigma
\rVert_{2}=o_{P}(1)$.  Because $\Sigma$ is positive-definite
(Condition~\ref{cond:cov_conv}), it is invertible and
$\lVert \hat{\Sigma}^{-1}\rVert \rightarrow_{P} \lVert
{\Sigma}^{-1}\rVert$. Applying sub-multiplicativity of the spectral
norm to the algebraic identity $\hat{\Sigma}^{-1} - \Sigma^{-1} =
\hat{\Sigma}^{-1}(\hat{\Sigma} - \Sigma)\Sigma^{-1}$, 
\begin{align*}
  \lVert\hat{\Sigma}^{-1} - \Sigma^{-1}\rVert_{2} \leq &
                                                         \lVert \hat{\Sigma}^{-1}\rVert_{2}
                                                         \lVert \hat{\Sigma} - \Sigma\rVert_{2}
                                                         \lVert
                                                         {\Sigma}^{-1}\rVert_{2}\\
  & = O_{P}(1) o_{P}(1) O(1) = o_{P}(1).
\end{align*}
Combining this with $\lVert \suppInstPenVecHat -
\suppInstPenVec\rVert_2 = O_P(n^{-1/2})$ by
Condition~\ref{cond:rootn}, $\lVert \hat{\Sigma}^{-1}\suppInstPenVecHat -
\Sigma^{-1}\suppInstPenVecHat\rVert_{2}=o_{P}(1)O_{P}(1)=o_{P}(1)$.
Separately  $\lVert \Sigma^{-1}\suppInstPenVecHat -
\Sigma^{-1}\suppInstPenVec \rVert_{2} = O_P(1)
O_{P}(n^{-1/2})=O_{P}(n^{-1/2})$. 
Thus,  $\lVert \hat{\Sigma}^{-1}\suppInstPenVecHat -
\Sigma^{-1}\suppInstPenVec\rVert_2 = o_P(1)$.  Now $\hat\omega = [\sum_j (\hat\Sigma^{-1}\suppInstPenVecHat)_{+_j}]^{-1}(\hat{\Sigma}^{-1}\suppInstPenVecHat)_+$,
and similarly $\omega = [\sum_j (\Sigma^{-1}\suppInstPenVec)_{+_j}]^{-1}\Sigma^{-1}\suppInstPenVec$;  
through an application of the Continuous Mapping Theorem, $\lVert \hat{\Sigma}^{-1}\suppInstPenVecHat -
\Sigma^{-1}\suppInstPenVec\rVert_2 = o_P(1)$ entails that the
normalizing constant in the definition of $\hat\omega$ converges to
the one in that of $\omega$. As a result, $\lVert \hat\omega -
\omega\rVert_{2} \rightarrow_{P} 0$.

We adopt a similar argument for the denominator.
$\lVert \hat\omega' \hat\Sigma \hat\omega - \hat\omega' \Sigma \hat\omega \rVert_2 = O_P(1)o_P(1)O_P(1)=o_P(1)$ and 
$\lVert \hat\omega' \Sigma \hat\omega - \omega' \Sigma \omega \rVert_2 = o_P(1)O_P(1)o_P(1)= o_P(1)$.
Thus, $|\hat{\omega}' \hat{\Sigma}\hat{\omega} -
\omega'\Sigma\omega| \rightarrow_{P} 0$, i.e. in 
\eqref{eq:t_cp_rootn} below the left denominator
 converges to the denominator at the right, a positive constant: 
\begin{equation}
  \label{eq:t_cp_rootn}
     \frac{\sqrt{n}(\sum_{g,t} \hat{\omega}_{gt}\hat{\Delta}_{gt} - \sum_{g,t}\hat{\omega}_{gt}\delta_{0gt})}{(\hat{\omega}'n\widehat{\Cov}(\hat{\Delta})\hat{\omega})^{1/2}} - \frac{\sqrt{n}(\sum_{g,t}\omega_{gt}\hat{\Delta}_{gt} - \sum_{g,t} \omega_{gt}\delta_{0gt})}{(\omega'n{\Cov}({\hat\Delta})\omega)^{1/2}}.
\end{equation}
Noting that \eqref{eq:t_cp_rootn} is equivalent to  the left-hand side
of \eqref{eq:t_cp} in the statement of the
Proposition,  
we just need to show that $\sqrt{n}\big[\sum_{g,t}\hat{\omega}_{gt}\hat{\Delta}_{gt} - \sum_{g,t}\hat{\omega}_{gt}\delta_{0gt} - \sum_{g,t}\omega_{gt}\hat{\Delta}_{gt} + \sum_{g,t}\omega_{gt}\delta_{0gt}\big] \rightarrow_P 0$, which is equivalent to showing  $\sqrt{n}\big[\sum_{g,t}(\hat{\omega}_{gt} - \omega_{gt})(\hat{\Delta}_{gt} - \delta_{0gt})\big] \rightarrow_P 0$. We have already demonstrated $\lVert\hat{\omega} - \omega\rVert_2 = o_P(1)$ and under the null distribution, $\lVert\hat{\Delta} - \delta_0\rVert_2 = O_P(n^{-1/2})$ through another application of the Weak Law of Large Numbers. Thus, $n^{1/2}[(\hat{\omega} - \omega)(\hat{\Delta} - \delta_0)] = o_P(1)$. 

\end{appendices}

\end{document}